\documentclass{article}

\usepackage{arxiv}
\usepackage{amsmath}
\usepackage{amsthm}
\usepackage{amsfonts}
\usepackage{amssymb}
\usepackage{cite} 
\usepackage{url} 
\usepackage{stackengine}
\usepackage{caption}
\usepackage{subcaption}
\usepackage[linesnumbered,ruled,vlined]{algorithm2e}
\usepackage{tikz}
\usetikzlibrary{automata, positioning, arrows, trees}

\tikzset{%
    ->,  
    >=stealth, 
    node distance=2cm, 
    every state/.style={thick, fill=gray!10}, 
    initial text=$ $, 
    }

\newcommand{\cla}[1]{\left[#1\right]}
\newcommand{\qua}[1]{\left<#1\right>}

\newtheorem{prop}{Proposition}
\newtheorem{coro}{Corollary}

\newcommand{\ra}{\ensuremath{\rightarrow}}
\newcommand{\indist}{\ensuremath{\equiv}}
\newcommand{\tequal}{\ensuremath{\approx}}
\newcommand{\proba}[1]{\ensuremath{\func_{#1}}}
\newcommand{\Nat}{\ensuremath{\mathbb{N}}}
\newcommand{\Real}{\ensuremath{\mathbb{R}}}

\newcommand{\Aut}{\ensuremath{A}}
\newcommand{\AutB}{\ensuremath{B}}

\newcommand{\quanti}{\ensuremath{I}}
\newcommand{\probP}{\ensuremath{P}}
\newcommand{\Sta}{\ensuremath{Q}}

\newcommand{\NTree}{\ensuremath{T}}

\newcommand{\func}{\ensuremath{f}}
\newcommand{\sta}{\ensuremath{q}}
\newcommand{\wor}{\ensuremath{s}}
\newcommand{\tol}{\ensuremath{t}}

\newcommand{\Symb}{\ensuremath{\Sigma}}
\newcommand{\Psimplex}{\ensuremath{\Delta}}

\newcommand{\symb}{\ensuremath{\sigma}}
\newcommand{\policy}{\ensuremath{\pi}}
\newcommand{\tra}{\ensuremath{\tau}}
\newcommand{\emptyW}{\ensuremath{\lambda}}
\newcommand{\quantp}{\ensuremath{\kappa}}
\newcommand{\pdist}{\ensuremath{\delta}}
\newcommand{\ce}{\ensuremath{\gamma}}

\newcommand{\qindist}[1]{\ensuremath{#1{/_{\scriptscriptstyle\indist}}}}
\newcommand{\qindistp}[1]{\ensuremath{#1{/_{\scriptscriptstyle\indist_\quantp}}}}

\newcommand{\terminal}{\ensuremath{\$}}
\newcommand{\SymbT}{\ensuremath{\Symb_\terminal}}
\newcommand{\staI}{\ensuremath{\sta_{\mathrm{in}}}}
\newcommand{\ProbT}{\ensuremath{\Psimplex(\SymbT)}}
\newcommand{\Words}{\ensuremath{\Symb^\ast}}
\newcommand{\policyW}{\ensuremath{\policy^\ast}}
\newcommand{\policyL}{\ensuremath{\policy^\ell}}
\newcommand{\traW}{\ensuremath{\tra^\ast}}
\newcommand{\probA}{\func_\Aut} 
\newcommand{\probB}{\func_\AutB}

\newcommand{\EQ}{\ensuremath{\mathbf{EQ}}}
\newcommand{\MQ}{\ensuremath{\mathbf{MQ}}}

\newcommand{\WLstar}{\ensuremath{\mathrm{WL}^\ast}}
\newcommand{\PLstar}{\ensuremath{\mathrm{L_{p}^\ast}}}

\newcommand{\QuaNT}{\ensuremath{\mathrm{QuaNT}}}

\newcommand{\DisS}{\ensuremath{Dis}}
\newcommand{\AccS}{\ensuremath{Acc}}
\newcommand{\Pref}{\ensuremath{Pre}}
\newcommand{\Suff}{\ensuremath{Suf}}

\title{Towards Efficient Active Learning of PDFA}
\renewcommand{\shorttitle}{Towards Efficient Active Learning of PDFA}

  \author{%
    {F. Mayr \hspace*{1.5ex} S. Yovine \hspace*{1.5ex} F. Pan} \\
    Facultad de Ingeniería\\
    Universidad ORT Uruguay\\
    Montevideo, Uruguay\\
    \texttt{mayr@ort.edu.uy}\\ 
    \texttt{yovine@ort.edu.uy}
   \And
    {N. Basset \hspace*{1.5ex} T. Dang}\\ 
    Verimag\\
    Université Grenoble-Alpes\\ 
    Grenoble, France\\
    \texttt{nicolas.basset1@univ-grenoble-alpes.fr}\\
    \texttt{thao.dang@univ-grenoble-alpes.fr}
   }

\renewcommand{\headeright}{Mayr F. et al.}

\begin{document}

\maketitle

\begin{abstract}
We propose a new active learning algorithm for PDFA based on three main aspects: a congruence over states which takes into account next-symbol probability distributions, a quantization that copes with differences in distributions, and an efficient tree-based data structure. Experiments showed significant performance gains with respect to reference implementations.
\end{abstract}


\section{Introduction}

We are interested in the problem of efficiently learning probabilistic deterministic finite automata (PDFA) in the context of the general active MAT-learning framework proposed in~\cite{Angluin:1987} where a \emph{learner} and a so-called \emph{minimum adequate teacher} interact by asking and responding questions, respectively. The learner's purpose is to unveil the hidden target automaton only known to the teacher. The latter allows the former to ask two kinds of questions, namely \emph{membership} (\MQ) and \emph{equivalence} (\EQ) queries.

\MQ\ seeks to discover the automaton's outcome for a particular string. 
Originally, the purpose of \MQ\ is to know whether a string belongs to the language to be learnt or, equivalently, whether it is accepted by the associated automaton. 
This name has been kept in later applications to other classes of automata.
In~\cite{bergadano1996, beimel_et_al}, the framework is used to prove learnability of multiplicity automata which compute functions mapping strings to elements of a \emph{field}. In this case, the name \MQ\ is retained even when the answer is the value of the computed function, which may not be Boolean. The same approach is followed in~\cite{weiss_WFA_learning} where \MQ\ is used to refer to the function that returns the probability of the last symbol in a string. 

\EQ\ provides means for determining to what extent the hypothesis automaton produced by the learner approximates the target one. In the case of \emph{exact} learning, this query checks whether they both produce the same outcome to \MQ\ for all strings. 
%
When \MQ\ are not binary, such as for PDFA, it could be useful to relax equality in order to learn \emph{similar} hypotheses that tolerate small discrepancies. For instance, in~\cite{clark_thollard, weiss_WFA_learning, tappler} equality is replaced by a \emph{similarity} relation.
However, that similarity relation is not an equivalence as it lacks transitivity. 

In the context of DFA, it is known that using appropriate data structures, such as trees, and efficiently processing counterexamples lead to significant gains in terms of computation time~\cite{introduction_to_computational_learning_theory, ttt}. To the best of our knowledge, these approaches have not yet been explored for active learning of PDFA.
The algorithms proposed in~\cite{katznacheev_et_al, beimel_et_al} rely on \MQ\ that compute the probability of a string and on an observation table to store the results.
In~\cite{tzeng}, \MQ\ return state distributions, that is, the probability that the target probabilistic automaton enters a state after reading an input string. This requires knowing the number of states of the target in advance. These works focused on theoretical results without implementations of them being publicly available.
In~\cite{kwiatkowska}, learning PDFA is a building block of an assume-guarantee framework for verification of probabilistic systems. \MQ\ asks for the probability of accepting a string and results are stored in an observation table. However, the overall goal is not to learn a PDFA equivalent to a hidden target but only an appropriate assumption for doing a compositional proof of correctness. To achieve this, the algorithm uses an \EQ\ that relies on language inclusion of PDFA and probabilistic model-checking. 
All these works rely on exact equality of \MQ\ outcomes. To deal with noise in distributions, \WLstar~\cite{weiss_WFA_learning}  proposes a non-equivalence similarity relation between probability distributions and develops an algorithm to learn similar hypotheses according to it. Note that \WLstar\ uses clustering to group responses to \MQ\ stored in the table. 

Based on these observations, we formulate a learning framework for PDFA where the teacher's answers to \MQ\ are probability distributions over the symbols a string could be continued with. These are called \emph{next-symbol} probability distributions. 
Besides, in the case the learner is allowed to produce hypotheses whose next-symbol probability distributions are approximations of the target ones, we propose to resort to quantization as an abstraction which allows us to define a coarser equivalence relation than the latter. This ensures the learnt automaton to be equivalent to the target one modulo this relation. 
Finally, we propose and implement \QuaNT, a learning algorithm for PDFA that uses an adaptation of the tree structure of~\cite{introduction_to_computational_learning_theory}. In order to assess its performance, we compared \QuaNT\ to a clustering-based algorithm that uses an observation table and a tolerance-based non-equivalence similarity relation, similar to \WLstar.
The experiments carried out showed that \QuaNT\ was orders of magnitude more efficient. 

The paper is organized as follows. Section~\ref{pdfa} gives core  definitions and results of interest. 
Section~\ref{sec:quant} presents our PDFA learning algorithm \QuaNT. Section~\ref{experiments} discusses experimental results. Section~\ref{sec:conclusions} summarizes the contributions.
\section{PDFA}\label{pdfa}

Let \Symb\ be a finite \emph{alphabet} and \SymbT\ to be the set $\Symb \cup \{\terminal\}$, where \terminal\ is a special \emph{terminal} symbol not in \Symb. 
\ProbT\ is probability simplex over \SymbT.
A 
PDFA \Aut\ over \Symb\ is a tuple $(\Sta, \staI, \policy, \tra)$, where 
\Sta\ is a finite set of \emph{states},
$\staI \in \Sta$ is an \emph{initial} state, 
$\policy : \Sta \ra \ProbT$ associates a probability distribution over \SymbT\ 
to each state, and 
$\tra : \Sta \times \Symb \ra \Sta$ is the \emph{transition} (total) function. Fig.~\ref{fig:pdfa_a}($a$-$b$) depicts an example.
Let $\probP_{\sta}$ be the probability distribution over \Words, such that $\probP_{\sta} = \probP(\wor | \sta)$ is the probability of $\wor\in\Words$ from state $\sta\in\Sta$, defined as:
\begin{align*}
    \probP(\emptyW | \sta) 
        = \policy(\sta)(\terminal),
        \emptyW \textit{ is the empty string,}
    &\quad
    \probP(\symb \wor | \sta) 
        = \policy(\sta)(\symb) \cdot \probP( \wor | \tra(\sta, \symb) ),
        \symb \in \Symb, \wor \in \Words
\end{align*}

A PDFA \Aut\ computes a function $\probA$ from \Words\ to $[0,1]$. For any string $\wor \in \Words$, $\probA(\wor) = \probP_{\staI}(\wor)$.
For instance, the PDFA in Fig.~\ref{fig:pdfa_a} maps the empty string \emptyW\ to 0 and every string $a^n$, $n \geq 1$, to $0.5^n$.
We define $\traW(\sta, \wor)$ to be the natural extension of \tra\ to strings, that is, the state reached by \Aut\ when going through \wor\ starting at state \sta:
\begin{align*}
    \traW(\sta, \emptyW) = \sta 
    &\qquad
    \traW(\sta, \symb \wor) = \traW( \tra(\sta, \symb), \wor )
\end{align*}
Similarly, we define $\policyW(\wor | \sta)$ to be the probability distribution of the state reached by \Aut\ when going through \wor\ from state \sta:
%
\begin{align*}
    \policyW(\wor | \sta) 
        &= \policy( \traW( \sta,\wor) )
\end{align*}
We denote by $\traW(s)$ and $\policyW(s)$ the state reached when going through \wor\ from the initial state \staI\  and the associated distribution, respectively.

\subsection{State equivalence}

We define the relation $\indist$ as follows: for every $\sta, \sta' \in \Sta$, $\sta\indist\sta'$ if for every $\wor\in\indist$, $\policyW(\wor|\sta) = \policyW(\wor|\sta')$. 
Clearly, \indist\ is an equivalence relation between states. 

\begin{prop}\label{prop:congruence}
$\forall\sta_1\indist\sta_2 \in \Sta$, 
    1) $\policy(\sta_1) = \policy(\sta_2)$, and
    2) $\forall \symb\in\Symb, \tra(\sta_1, \symb)\indist\tra(\sta_2, \symb)$.
That is, \indist\ is a congruence.
\end{prop}

It follows that \indist\ induces a \emph{quotient} PDFA $\qindist{\Aut} = ( \qindist{\Sta}, \cla{\staI}, \qindist{\policy}, \qindist{\tra} )$, where 
\qindist{\Sta} is the set of equivalence classes, $\cla{\sta}\in\qindist{\Sta}$ denotes the class of $\sta\in\Sta$, and for any $\sta\in\Sta$, 
the probability distribution is
$\qindist{\policy}(\cla{\sta}) = \policy(\sta)$, and
the transition function is
$\qindist{\tra}(\cla{\sta}, \symb) = \cla{\tra(\sta, \symb)}$ for any $\symb\in\Symb$.
\footnote{This quotient is different from~\cite{thollard-dupont-delahiguera} defined by a partition resulting from merging states according to a compatibility criterion which is not an equivalence relation.}

\begin{prop}\label{prop:quotient}
For every PDFA \Aut, \qindist{\Aut} computes the same function as \Aut. 
\end{prop}
\noindent
Therefore, \indist\ can be extended to PDFA. For every PDFA \Aut\ and \AutB, $\Aut\indist\AutB$ if their respective initial states $\staI^\Aut$ and $\staI^\AutB$ are equivalent, that is, $\staI^\Aut \indist \staI^\AutB$. 

\begin{coro}\label{coro:equiv}
$\Aut\indist\AutB$ implies $\probA = \probB$. The converse does not hold in general.
\end{coro}
\noindent


\subsection{Minimality}

A PDFA is \emph{minimal} if any other PDFA that computes the same function has no less states~\cite{beimel_et_al}. The simple PDFA in Fig.~\ref{fig:pdfa_a} is minimal since clearly the function cannot be computed by a PDFA with a single state.
Based on \indist\ we define a weaker notion of minimality as follows. A PDFA \Aut\ is said to be \emph{weakly minimal} if for every $\sta, \sta' \in \Sta$, $\sta \not\indist \sta'$.
By definition, for every PDFA \Aut, \qindist{\Aut} is weakly minimal.

\begin{prop}\label{prop:wm}
Every minimal PDFA \Aut\ is also weakly minimal. The converse does not hold in general.
\end{prop}

An example of Proposition~\ref{prop:wm} is the PDFA in Fig.~\ref{fig:pdfa_a}.
However, the converse is not true, that is, there are weakly minimal PDFA which are not minimal.
Consider for instance PDFA \Aut\ and \AutB\ in Fig.~\ref{fig:pdfa_A_B}. We have that $\probB((ab)^{n+1}) = (0.1 \cdot 0.2) (0.2 \cdot 0.1)^n \cdot 0.3 = (0.1 \cdot 0.2)^{n+1} \cdot 0.3 = \probA((ab)^{n+1})$, for $n \geq 0$. For any other string $\wor \neq (ab)^{n+1}$, $\probA(\wor) = \probB(\wor) = 0$. Hence, \Aut\ and \AutB\ compute the same function. Moreover, they are both weakly minimal because $\policyW(\emptyW|\sta) \neq \policyW(\emptyW|\sta')$ for every pair of states $\sta,\sta'\in\Sta$, for each one of the automata, respectively. But \AutB\ is certainly not minimal because it has more states than \Aut. This example also shows that $\probA = \probB$ does not entail $\Aut\indist\AutB$.

\subsection{Similarity of distributions}

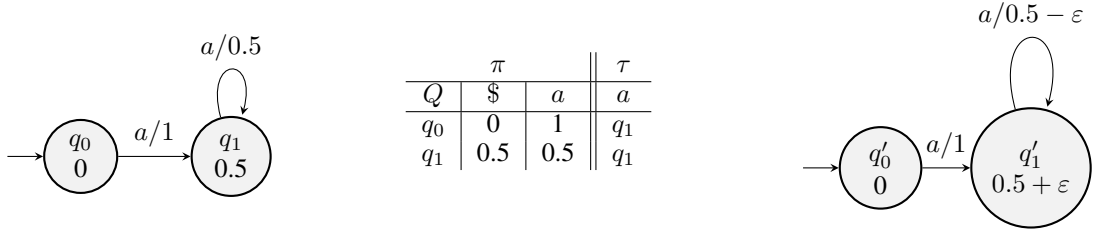
\begin{figure}[htbp]
\centering
    \begin{subfigure}[c]{0.33\textwidth}%
    \begin{tikzpicture}
        \node[state, initial] (q0) {\stackanchor{$q_0$}{0}};
        \node[state, right of=q0] (q1) {\stackanchor{$q_1$}{0.5}};
        \draw   
            (q0) edge[above] node{$a/1$} (q1)
            (q1) edge[loop above] node{$a/0.5$} (q1);
    \end{tikzpicture}
    \end{subfigure}
    \begin{subfigure}[c]{0.3\textwidth}%
    \begin{tabular}{c|c|c||c}
        \multicolumn{3}{c||}{\policy} & \tra \\ \hline
        \Sta    & \terminal\  & $a$ & $a$ \\ \hline
        $q_0$   &  0          & 1   & $q_1$ \\
        $q_1$   &  0.5        & 0.5 & $q_1$ \\
    \end{tabular}
    \end{subfigure}
    \begin{subfigure}[c]{0.33\textwidth}%
        \begin{tikzpicture}
        \node[state, initial] (q0) {\stackanchor{$q'_0$}{0}};
        \node[state, right of=q0] (q1) {\stackanchor{$q'_1$}{$0.5+\varepsilon$}};
        \draw   
            (q0) edge[above] node{$a/1$} (q1)
            (q1) edge[loop above] node{$a/0.5-\varepsilon$} (q1);
    \end{tikzpicture}
    \end{subfigure}
    \caption{($a$-$b$) PDFA over $\Symb = \{a\}$ with $\staI = q_0$. ($c$) PDFA $\Aut_{\varepsilon}$.}
    \label{fig:pdfa_a}
\end{figure}

The equivalence defined above is not robust in the sense that two states whose probability distributions differ very slightly are not equivalent. 
Consider the PDFA $\Aut_\varepsilon$ (c) in Fig.~\ref{fig:pdfa_a}, where $\varepsilon\in(0,0.5]$. It maps every string $a^n$, $n \geq 1$, to $(0.5-\varepsilon)^{n-1} (0.5+\varepsilon)$. This PDFA is not equivalent to PDFA \Aut\ in Fig.~\ref{fig:pdfa_a}. 
However, \proba{\Aut_\varepsilon} tends to $\probA$ as $\varepsilon$ tends to $0$. Moreover, $\policy(q'_1)$ tends to $\policy(q_1)$.
%
%
Previous works have addressed this issue by introducing a \emph{tolerance} parameter $\tol$. 
%
%
In~\cite{weiss_WFA_learning}, states are compared using the probability of the last symbol of a non-empty string defined as:
\begin{align}\label{eq:policyL}
   \policyL_{\sta}(\wor\symb) = \policyW(\wor|\sta)(\symb)
   &\;\mathrm{where}\; \symb \in \SymbT, \wor \in \Words, \sta \in \Sta
\end{align}
It is said that $\sta, \sta'$ are $\tol$-equal, for $\tol\in [0,1]$, denoted $\sta\tequal_\tol\sta'$, if
$L_{\infty}(\policyL_{\sta}, \policyL_{\sta'}) \leq \tol$ where $L_{\infty}(v,v')=\max_{x} |v(x)-v'(x)|$.
PDFA \Aut\ and \AutB\ are \tol-equal, denoted $\Aut\tequal_\tol\AutB$, if  $\staI^\Aut\tequal_\tol\staI^\AutB$.
%
However, this approach does not lead to an equivalence relation between states. That is why, to cope with small perturbations in probabilities while preserving equivalence between states, we propose to resort to a quantization defined over \policyW as follows.

Let $\quantp\in\Nat$, $\quantp\geq 1$, be a \emph{quantization} parameter. For $n\in\Nat$, $0\leq n<\quantp-1$, we define $\quanti_\quantp^n$ to be the left-closed right-open interval $\left[ n \quantp^{-1}, (n+1) \quantp^{-1} \right)$, and for $n=\quantp-1$, the closed interval $\left[ n \quantp^{-1}, 1 \right]$. Now, for every real number $x\in[0,1]$, we define $\qua{x}_\quantp = \quanti_\quantp^n$ such that $x \in \quanti_\quantp^n$. 
For instance, for $\quantp = 2$, we have the quantization $[0, 0.5), [0.5, 1]$. 
Two numbers $x, y \in\Real$ are $\quantp$-equivalent, denoted $x =_\quantp y$ if $\qua{x}_\quantp = \qua{y}_\quantp$. This definition extends naturally to \ProbT. We denote $\qua{\ProbT}_\quantp$ the partition of \ProbT\ induced by \quantp.
For $\pdist\in\ProbT$, $\qua{\pdist}_\quantp$ is called the \emph{quantization vector} of $\pdist$. For example, $\qua{(0.1, 0.3, 0.6)}_2 = (\quanti^0_2, \quanti^0_2, \quanti^1_2)$.

Being $=_\quantp$ an equivalence relation, we can define a quantized version of $\indist$ as follows: $\sta\indist_\quantp\sta'$ if for every $\wor\in\Words$, $\policyW(\wor|\sta) =_\quantp \policyW(\wor|\sta')$. This induces a unique quotient over states $\qindistp{\Sta}$.
Moreover, $\indist_\quantp$ can be extended to PDFA: for every PDFA \Aut\ and \AutB, $\Aut\indist_\quantp\AutB$ if $\staI^\Aut \indist_\quantp \staI^\AutB$.
This allows extending the notion of weakly minimality defined over \indist\ to $\indist_\quantp$: \Aut\ is said to be \quantp-\emph{weakly minimal} if for every $\sta, \sta' \in \Sta$, $\sta \not\indist_\quantp \sta'$.
However, unlike \indist, its quantized version $\indist_\quantp$ does not induce a unique quotient PDFA because $\qindistp{\policy}(\cla{\sta}_\quantp)$ can be \emph{any} distribution in $\qua{\policy(\sta)}_\quantp$. 
Therefore, we define $\qindistp{\Aut} = \{ \AutB \mid \Aut\indist_\quantp\AutB  \land |\qindistp{\Sta^\Aut}| = |\Sta^\AutB| \}$, that is, the set of \quantp-weakly minimal PDFA which are \quantp-equivalent to \Aut.

To illustrate quantization, consider the PDFA \Aut\ (a) and $\Aut_{\varepsilon}$ (c) in Fig.~\ref{fig:pdfa_a}. Suppose $\varepsilon\in(0,0.1)$. For $\quantp = 5$, we have the quantization $[0, 0.2), \ldots, [0.8, 1]$. Then, $\qua{\policy(q_1)(a)}_5 = \quanti^2_5 = [0.4,0.6) = \qua{\policy(q'_1)(a)}_5$, since $0.4 < 0.5-\varepsilon < 0.5$, and $\qua{\policy(q_1)(\terminal)}_5 = \quanti^2_5 = [0.4,0.6) = \qua{\policy(q'_1)(\terminal)}_5$, since $0.5 < 0.5+\varepsilon < 0.6$.
Thus, we have that $q_i \indist_5 q'_i$. Hence, $\Aut \indist_5 \Aut_\varepsilon$. 

\begin{prop}\label{prop:quantp_vs_tol}
For every $\sta, \sta'$, if $\sta\indist_\quantp\sta'$ then $\sta\tequal_{\quantp^{-1}}\sta'$. 
\end{prop}
%
\section{PDFA Learning: QuaNT}\label{sec:quant}

\QuaNT\ is a learning algorithm that constructs a PDFA by interacting with a teacher which makes use of oracles $\MQ_{\QuaNT}$\ and $\EQ_{\QuaNT}$. It has three major differences with \WLstar.
%
First, $\MQ_{\QuaNT}$ returns the next-symbol probability simplex of a string, that is $\MQ_{\QuaNT}(\wor) = \policyW(\wor)$ for $\wor\in\Words$; while $\MQ_{\WLstar}(\wor)$ returns $\policyL_{\staI}(s)$, where $\policyL$ is defined in (\ref{eq:policyL}).
Second, it relies on quantization rather than on tolerance for comparing probability distributions. Since our quantization method induces a congruence, there is no need for clustering states. 
Third, along the lines of~\cite{introduction_to_computational_learning_theory} and ~\cite{ttt} algorithms, \QuaNT\ builds a \emph{classification} tree instead of a table. Similarly to these algorithms, tree leafs are PDFA states identified by so-called \emph{access} strings and inner nodes are \emph{distinguishing} suffixes. Nevertheless, in \QuaNT\, the tree is not binary but $n$-ary, where $n$ is the number of different classes  which \ProbT\ is partitioned into, and tree arcs and leafs are labelled with elements of $\qua{\ProbT}_\quantp$ and \ProbT, respectively.

\paragraph{N-ary classification tree}
Given $\wor \in \Words$, we denote $\cla{\wor}_\quantp$ the class of the state reached by traversing $\wor$ in the target PDFA \Aut, i.e., $\cla{\traW(\wor)}_\quantp$. Hereinafter, we omit the quantization parameter \quantp\ when clear from context. 
The tree $\NTree_{\AccS,\DisS}$ maintains a set \AccS\ of access strings and a set \DisS\ of distinguishing strings where for every pair of distinct strings $\wor, \wor' \in \AccS$:
    (1) $\cla{\wor}\neq\cla{\wor'}$, (that is, $\traW(\wor) \not \indist_{\quantp} \traW(\wor')$), and
    (2) $\exists d \in \DisS$ s.t. $\policyW(\wor d) \neq_\quantp\policyW(\wor' d)$.
%
%
%
The distinguishing string that labels the root of the tree is always \emptyW\ which is also an access string, so that the initial state of the PDFA can be accessed.

\begin{prop}\label{prop:tree-size}
The number of leaves of \NTree\ is at most $|\qindistp{\Sta}|$.
\end{prop}

\begin{figure}[htbp]
\centering
    \begin{subfigure}[c]{0.45\textwidth}
    \begin{tikzpicture}
    [level distance=2.5cm,
    level 1/.style={sibling distance=1.5cm},
    level 2/.style={sibling distance=2.2cm}]
    \node {$\emptyW$}
    child { 
        node {\stackanchor{$\emptyW$}{$(0, 0.5, 0.5)$}}
        edge from parent node[left=0.15] {$(0, 5, 5)$}
    }
    child { 
        node {$1$}
        child {
            node {\stackanchor{$10$}{$(0.1, 0.3, 0.6)$}}
            edge from parent node[left=0.15] {$(1, 6, 3)$}
        }
        child {
            node {\stackanchor{$0$}{$(0.1, 0.3, 0.6)$}}
            edge from parent node[right=0.15] {$(1, 3, 6)$}
        }
        edge from parent node {$(1, 3, 6)$}
    }
    child { 
        node {\stackanchor{$1$}{$(0.1, 0.6, 0.3)$}}
        edge from parent node[right=0.15] {$(1, 6, 3)$}
    };
    \end{tikzpicture}
    \end{subfigure}
    \begin{subfigure}[c]{0.45\textwidth}
    \begin{tikzpicture}
        [node distance=2.7cm]
        \node[state, initial, initial where=above] (q0) {\stackanchor{$q_\emptyW$}{$0$}};
        \node[state, below left of=q0] (q1) {\stackanchor{$q_1$}{$0.1$}};
        \node[state, below right of=q0] (q2) {\stackanchor{$q_0$}{$0.1$}};
        \node[state, below right of=q1] (q3) {\stackanchor{$q_{10}$}{$0.1$}};
        \draw   
            (q0) edge[left, bend right] node{$1/0.5$} (q1)
            (q0) edge[right, bend left] node{$0/0.5$} (q2)
            (q1) edge[loop left] node{$1/0.3$} (q1)
            (q1) edge[left, bend right] node{$0/0.6$} (q3)
            (q2) edge[loop left] node{$0/0.3$} (q2)
            (q2) edge[right, bend left] node{$1/0.6$} (q3)
            (q3) edge[left, bend right] node{$1/0.6$} (q1)
            (q3) edge[right, bend left] node{$0/0.3$} (q2);
    \end{tikzpicture}
    \end{subfigure}
    \caption{An n-ary classification tree (left), and the corresponding PDFA (right)}
    \label{fig:n_ary_tree_example}
\end{figure}
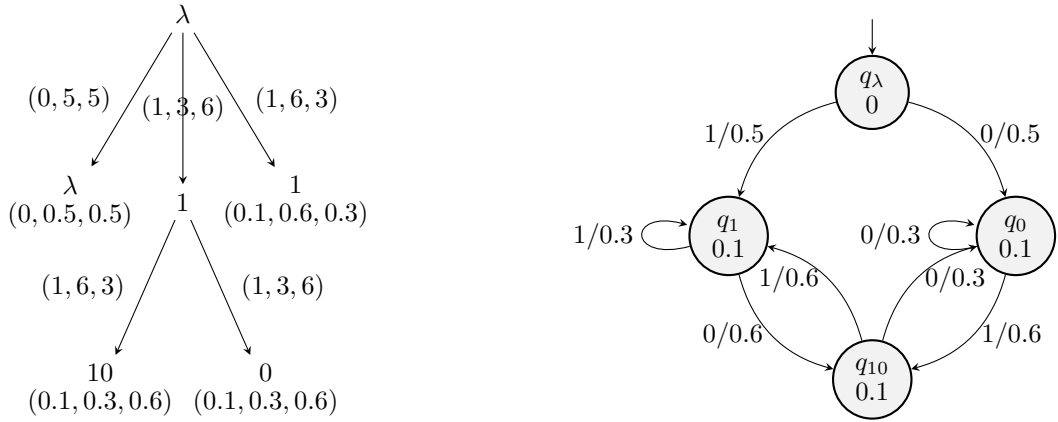

Fig.~\ref{fig:n_ary_tree_example} (left) shows an example of $n$-ary classification tree. The quantization parameter is $\quantp = 10$. Tree leafs correspond to the states of the PDFA shown on the right, identified with their associated access strings: $\AccS = \{\emptyW, 0, 1, 10\}$. Every leaf is labeled with a vector corresponding to the probability distribution of the state, where the first element is the probability of \terminal, the second element is the probability of $0$, and the last element is the probability of $1$. For example: $\policy(q_\emptyW) = (0, 0.5, 0.5)$. 
Tree arcs are labeled with quantization vectors. To simplify the visualization, only partition indexes are shown. For instance, $(0, 5, 5)$ corresponds to the quantization vector $(\quanti_{10}^0, \quanti_{10}^5, \quanti_{10}^5)$.
The root \emptyW\ of the tree has an arc for each one of the classes in which quantization partitions the set of probability distributions of the states of the PDFA. In the example, there are three, namely quantization vectors $(0,5,5)$, $(1,6,3)$, and $(1,3,6)$, corresponding to distributions $(0,0.5,0.5)$, $(0.1,0.6,0.3)$, and $(0.1,0.3,0.6)$, respectively. The tree explains that states $q_\emptyW$ and $q_1$ are not equivalent, i.e., $\cla{\emptyW} = q_\emptyW \not\indist q_1 = \cla{1}$, because $\policyW(\emptyW|q_\emptyW) = \policy(q_\emptyW) \neq \policy(q_1) = \policyW(\emptyW|q_1)$, and they are distinguished by $\emptyW\in\DisS$. They are also not equivalent to $q_{0}$ and $q_{10}$ for the same reason. These two states, which have the same probability distribution, are indeed distinguished by the string $1\in\DisS$ because $\policyW(1|q_{0}) = \policy(q_{10}) \neq \policy(q_{1}) = \policyW(1|q_{10})$.

\paragraph{Finding the class of a state (sifting)}
Given a string $\wor'\not\in \AccS$, the tree allows to efficiently determine its class $\cla{\wor'}$. That means either finding an access string $\wor\in\AccS$ such that $\cla{\wor'}=\cla{\wor}$ or creating a new class by adding  $\wor'$ to \AccS. To this end, we define the \emph{sifting} operation as follows. Sift starts at the root of \NTree. Let $d\in\DisS$ be the distinguishing string at the current node of the tree. In this case, we perform a membership query to get $\qua{\MQ_{\QuaNT}(\wor'd)}$ and then we descend to the subtree labeled with such quantization vector. Sift continues in this manner until a leaf $\wor\in\AccS$ is reached, in which case, $\cla{\wor}=\cla{\wor'}$. If there is no arc labeled with the same quantization vector, we have discovered a new class and must update the tree (\emph{sift-update}) by adding $\wor'$ to $\AccS$ with probability $\MQ_{\QuaNT}(\wor')$ and a new arc $(d, \qua{\MQ_{\QuaNT}(\wor'd)}, \wor')$.
Sifting can be efficiently implemented because the number of membership queries is bounded by the depth of \NTree, and finding quantization vectors and asking membership queries can be cached to minimize the number of vector comparisons and queries to the target system, respectively.

\paragraph{Building a tentative hypothesis $\widehat{\Aut}$}
Given a tree \NTree, it is easy to construct $\widehat{\Aut}$ using sift. Each state of $\widehat{\Aut}$ is uniquely identified with an access string in \AccS. For each state $q_\wor$, $\wor \in \AccS$, and symbol $\symb \in \Symb$, $\tra(q_\wor, \symb) = q_{\wor'}$, where $\wor' = \mathrm{sift}(\wor\symb)$, and $\policy(q_\wor)$ is the probability distribution associated with the leaf node \wor. In the case that sift generates an update, the building process is restarted.
Because of Prop.~\ref{prop:tree-size}, this eventually terminates.
Moreover, by construction, $\widehat{\Aut}$ is a PDFA.

\paragraph{Equivalence queries}
For $\EQ_{\QuaNT}$ we use an adaptation of Hopcroft-Karp where states are compared as follows: $\qua{\policy(\widehat{\sta})}_\quantp = \qua{\policy(\sta)}_\quantp$, where \sta\ is a state of the target PDFA \Aut\ and $\widehat{\sta}$ is a state of the hypothesis PDFA $\widehat{\Aut}$. 

\paragraph{Processing a counterexample}
Let \ce\ be a counterexample returned by $\EQ_{\QuaNT}$, that is $\policyW_{\Aut}(\ce) \neq_\quantp \policyW_{\widehat{\Aut}}(\ce)$. Let $\ce_{i}$ be the $i$-th symbol of $\ce$, 
$\ce[i]$ be the prefix of $\ce$ of length $i$, i.e., $\ce[i] = \ce_{1}...\ce_{i}$, 
$\wor_i = \mathrm{sift}(\ce[i])$, and 
$\widehat{\wor}_i$ the string associated to the state $\traW_{\widehat{\Aut}}(\ce[i])$.
Let $1\leq j \leq |\ce|$ be the first index such that $\widehat{\wor}_j \neq \wor_j$. This means that $\cla{\widehat{\wor}_j} \neq \cla{\wor_j}$ and
$\widehat{\wor}_{j-1} = \wor_{j-1}$, but the states reached by $\widehat{\Aut}$ and \Aut\ after traversing $\ce[j-1]$ are not equivalent since when continued with $\ce_j$ they reach non-equivalent states. That is, $\cla{\wor_{j-1}} \neq \cla{\ce[j-1]}$. Therefore, \NTree\ has to be updated by adding a new leaf node $\ce[j-1]$ representing the newly discovered class. Then, $\ce[j-1]$ is added to $\AccS$.
Now, let $d\in\DisS$ be the least common ancestor in \NTree\ of $\widehat{\wor}_j$ and $\wor_j$. Then $\ce_j d$ is a distinguishing string for $\wor_{j-1}$ and $\ce[j-1]$.
In terms of tree operations, the leaf $\wor_{j-1}$ is replaced by an inner node $\ce_{j}d$ and two children, namely $\wor_{j-1}$ and $\ce[j-1]$.\\

\begin{algorithm}
    \SetKwInOut{Input}{Input}
    \SetKwInOut{Output}{Output}
    \SetKwInOut{Parameter}{Parameter}
    
    \Parameter{Quantization Parameter $\kappa$}
    \Output{PDFA $\widehat{\Aut}$}
        
    $\widehat{\Aut} \leftarrow \mathrm{BuildSingleStatePDFA}(\quantp)$\;
    $\ce \leftarrow \EQ_{\QuaNT}(\widehat{\Aut}, \quantp)$\;
    
    \If{$\ce=\bot$}{
        \Return $\widehat{\Aut}$\;
    }
    $\NTree \leftarrow  \mathrm{InitializeTree}(\ce, \quantp)$\;
    
    \While{$\ce\neq\bot$}{
        $\widehat{\Aut} \leftarrow \mathrm{BuildAutomaton}(\NTree)$\;
        $\ce \leftarrow \EQ_{\QuaNT}(\widehat{\Aut}, \quantp)$\;
        
        \If{$\ce\neq\bot$}{
            $\NTree \leftarrow \mathrm{UpdateTree}(\NTree, \ce, \quantp)$\;
        }
    }
    \Return $\widehat{\Aut}$\;
    \caption{\QuaNT}
    \label{alg:quant}
\end{algorithm}

\paragraph{Complete algorithm}
Algorithm~\ref{alg:quant} shows \QuaNT\ pseudocode.
The algorithm begins by executing BuildSingleStatePDFA
which creates an initial hypothesis $\widehat{\Aut}$, with a single state $q_\emptyW$ with a loop for each symbol and executes $\MQ_{\QuaNT}(\emptyW)$ to get the probability distribution for the state.
Then, it calls $\EQ_{\QuaNT}(\widehat{\Aut}, \quantp)$, which either returns $\bot$ and terminates or a counterexample \ce\ enabling the initialization of the tree. The first tree \NTree\ has a root labeled with the distinguishing string \emptyW\ and two children, one with the access string \emptyW\ and the other with the counterexample \ce. 
%
Once \NTree\ is initialized, the main loop of the algorithm begins. It consists in using \NTree\ to build a PDFA $\widehat{\Aut}$, then using $\EQ_{\QuaNT}$\ to compare it with the target PDFA \Aut. If a counterexample is returned, \NTree\ is updated, resulting in new states being discovered. Otherwise it means all states in $\qindistp{\Sta}$ have been found which implies $\widehat{\Aut} \indist_\quantp \Aut$.

\begin{prop}\label{prop:termination}
For any PDFA \Aut, \QuaNT\ terminates and computes a PDFA $\widehat{\Aut} \in \qindistp{\Aut}$.
\end{prop}


%
\section{Experiments}\label{experiments}
In this section we present the results of the experiments carried out to compare \QuaNT\ with a clustering-based algorithm that uses an observation table and a tolerance-based non-equivalence similarity relation, similar to \WLstar, that we call \PLstar\ (see Appendix~\ref{apd:plstar}). 
We compared the learning algorithms on randomly generated PDFA. 
The generation technique works in two steps. First, it constructs random DFA over \Symb. Second, DFA are transformed into PDFA by assigning a probability distribution over \SymbT\ to every state.
%
The first step uses the method described in \cite{nicaud} based on results from \cite{distribution_of_states}. Let $n$ be the desired number of reachable states of a DFA, called its \emph{nominal} size. The method consists in randomly generating DFA of a total of $n\cdot m\cdot \rho_{m}^{-1}$ possibly unreachable states, for $m = |\Symb|$, where  $\rho_{m} = m - W_{0}\cdot m\cdot e^{-m}$ and $W_{0}$ is the Lambert-W function, and then computing its accessible part by a depth-first traversal.
It is important to remark that this method does not guarantee the \emph{actual} size of the accessible part to be exactly $n$, but to be normally distributed around $n$. To obtain exactly $n$ accessible states, the method could be repeated using a rejection algorithm. However, in practice, this proved to be very inefficient, being almost impossible to generate DFA of accessible size bigger than 100 in reasonable time. 
All experiments threw perfect scores for all computed metrics (word error rate, normalized discounted cumulative gain, log probability error~\cite{weiss_WFA_learning}) for all algorithms on the same test set of strings. Therefore, the analyses of the experimental results are mainly focused on execution time and structure size. For ease of comparison, figures show trend lines.

\subsection{Experiment 1}\label{exp:1}
\begin{figure}[htbp]
\centering
    \begin{subfigure}[c]{0.45\textwidth}
    \includegraphics[width=\textwidth]{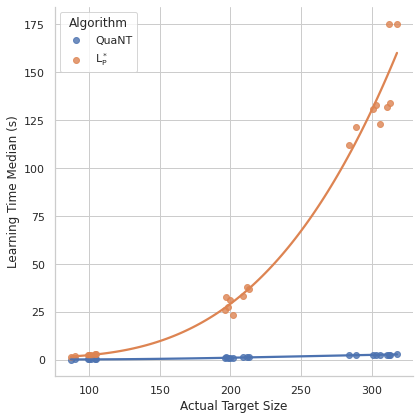}
        \caption{Execution time.}
        \label{fig:exp1_time}
    \end{subfigure}
    \begin{subfigure}[c]{0.45\textwidth}
    \includegraphics[width=\textwidth]{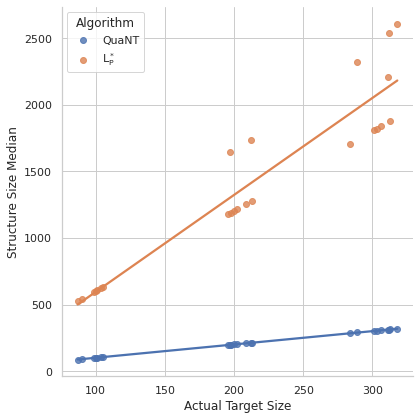}
        \caption{Structure size.}
        \label{fig:exp1_str}
    \end{subfigure}
    \caption{Experiment 1}
    \label{fig:exp1}
\end{figure}

\begin{figure}[htbp]
\centering
    \begin{subfigure}[c]{0.45\textwidth}
    \includegraphics[width=\textwidth]{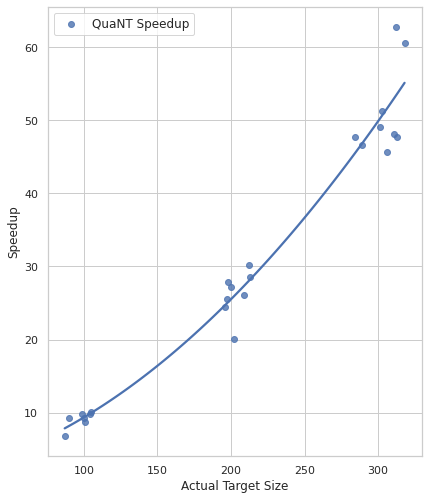}
    \caption{Experiment 1 (Speedup).}
    \label{fig:exp1_speedup}
    \end{subfigure}
    \begin{subfigure}[c]{0.45\textwidth}
    \includegraphics[width=\textwidth]{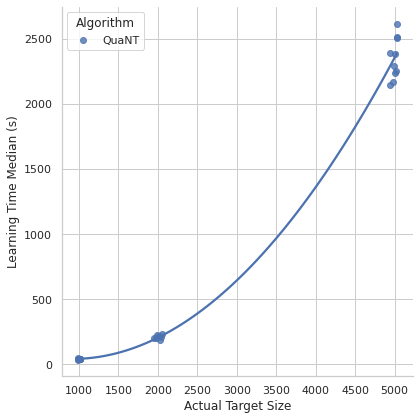}
    \caption{Experiment 4.}
    \label{fig:exp4}
    \end{subfigure}
    \caption{Experiments 1 (Speedup) and 4.}
    \label{fig:exp1_4}
\end{figure}

In this experiment we compared \QuaNT\ and \PLstar. For this, 10 random PDFA over a binary alphabet ($m=2$) of nominal sizes $n = 100, 200, 300$ were generated, and each algorithm was run 10 times for each PDFA.
For \QuaNT\  $\quantp = 1000$, and for \PLstar,
$\tol = \quantp^{-1}$ (Prop~\ref{prop:quantp_vs_tol}). 
Fig.~\ref{fig:exp1_time} shows learning time medians for every actual size. Notably, \PLstar\ 
execution time grows much faster than \QuaNT's. Indeed, \QuaNT\ achieves a speedup of approximately $0.2n$, reaching around 60x for the biggest PDFA (see Fig.~\ref{fig:exp1_speedup}). 
This experiment also showed that the size of \PLstar's 
observation table grows bigger than \QuaNT's tree which partly explains the gains in execution time (Fig.~\ref{fig:exp1_str}).



\subsection{Experiment 2}\label{exp:2}
\begin{figure}[htbp]
\centering
    \begin{subfigure}[c]{0.45\textwidth}
    \includegraphics[width=\textwidth]{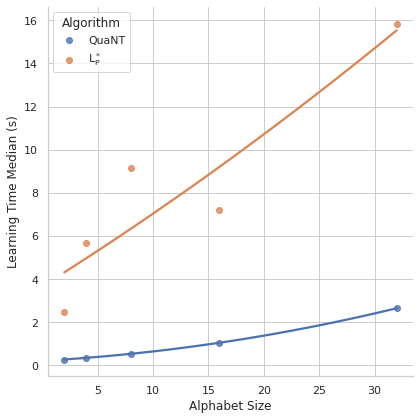}
    \caption{Experiment 2.}
    \label{fig:exp2}
    \end{subfigure}
    \begin{subfigure}[c]{0.45\textwidth}
    \includegraphics[width=\textwidth]{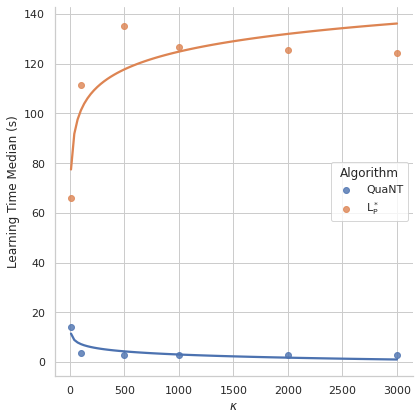}
    \caption{Experiment 3.}
    \label{fig:exp3}
    \end{subfigure}
    \caption{Experiments 2 and 3.}
    \label{fig:exp2_3}
\end{figure}

In this experiment 10 random PDFA of nominal size $n=100$ were generated for alphabet size $m=2,4,8,16,32$. We compared \QuaNT\ and \PLstar\  with $\quantp = 1000$ and $\tol = \quantp^{-1}$. Each algorithm was run 10 times for each PDFA.
Fig.~\ref{fig:exp2} shows the learning time medians for every alphabet size. As it can be seen \PLstar\ seems to be more sensitive to the growth in the alphabet size.

\subsection{Experiment 3}\label{exp:3}
In this experiment, we compare the algorithms for different values of tolerance and quantization parameter: $\quantp=10, 100, 500, 1000, 2000, 3000$, with $\tol = \quantp^{-1}$.
For every parameter configuration 10 random PDFA of nominal size $n = 300$ and alphabet size $m = 2$ were generated, and each algorithm was run 10 times for each PDFA.
Fig.~\ref{fig:exp3} shows the median learning times. As it can be seen both algorithms appear to stabilize its execution time after some parameters sizes ($\quantp = 500$, $t= 1/1000$).
    
\subsection{Experiment 4}\label{exp:4}
Here, \QuaNT\ was evaluated on bigger nominal sizes $n = 1000, 2000, 5000$, fixing $\quantp = 1000$ and $m=2$.
For every parameter configuration, 10 random PDFA were generated and each algorithm was run 10 times for each PDFA.
Fig.~\ref{fig:exp4} shows median learning times. Clearly, \QuaNT\ still manages to learn PDFA from systems that are 
intractable for \PLstar. Assuming a linear speedup of $0.2 n$ from Experiment \ref{exp:1}, the learning time of a PDFA of size 5000 would be almost a month for \PLstar.


\subsection{Experiment 5}\label{exp:5}
In previous experiments it is noted that nearly all states in the randomly generated PDFA have distinct next symbol distributions, that is, most states are distinguished by $\emptyW$, thus producing shallow tree structures (depth 1 or 2). In order to analyze cases where states share next symbol distributions we parameterized the PDFA random generation by a number $d$ of distributions to use. The algorithm first randomly generates a set of $d$ distributions and then labels each state by uniformly picking one in this set. 

\begin{figure}[htbp]
\centering
    \begin{subfigure}[c]{0.45\textwidth}
    \includegraphics[width=\textwidth]{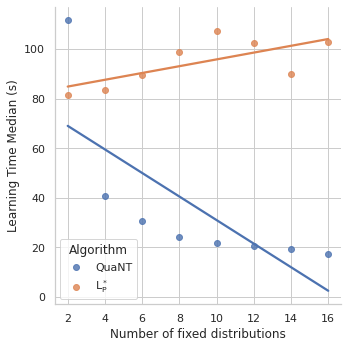}
    \caption{Execution time.}
    \label{fig:exp5_time}
    \end{subfigure}
    \begin{subfigure}[c]{0.45\textwidth}
    \includegraphics[width=\textwidth]{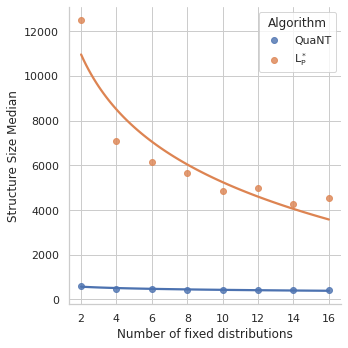}
    \caption{Structure size.}
    \label{fig:exp5_str}
    \end{subfigure}
    \caption{Experiment 5.}
    \label{fig:exp5}
\end{figure}

For this experiment, 10 random PDFA over a binary alphabet ($m = 2$) of nominal size $n = 300$ were generated for different values of $d$ (ranging from 2 to 16). Each algorithm was run 10 times for each PDFA. 
For \QuaNT,  $\quantp = 1000$, and for \PLstar,
$\tol = \quantp^{-1}$ (Prop~\ref{prop:quantp_vs_tol}). 
Fig.~\ref{fig:exp5_time} shows learning time medians for every $d$ value. Notably, \PLstar\ 
execution time still grows faster than \QuaNT's. The only case where \QuaNT\ achieves worse learning time than $\PLstar$ is for $d = 2$. Otherwise, \QuaNT\ significantly benefits from the increase in $d$. 
Regarding structure sizes, \PLstar's observation table grows bigger than previous experiments, being negatively affected by smaller values of $d$. However \QuaNT\ maintains similar sizes to those observed in experiment 1 (Fig.~\ref{fig:exp5_str}).
\section{Conclusions}\label{sec:conclusions}
We defined a robust notion of similarity of states in PDFA based on a congruence over states and a quantization of their probability distributions. This induces a precisely defined inductive bias for PDFA learning as a set of quantized weakly minimal PDFA.
Based on this, we developed a new PDFA active MAT-learning algorithm called \QuaNT\ which uses an n-ary tree to efficiently learn a PDFA in the hypothesis space.
In order to empirically assess the efficiency of \QuaNT, we presented an adaptation of \WLstar, namely \PLstar, which works with a non-equivalence similarity relation over distributions and an observation table. Algorithms were compared on a number of randomly generated PDFA. The experiments confirmed notable execution time gains achieved by \QuaNT.
%



\paragraph{Acknowledgments}\label{sec:aknowledgments}
Research reported in this article has been partially funded by the following grants: ANII-Agencia Nacional de Investigaci\'on e Innovaci\'on FMV\_1\_2019\_1\_155913, and the ANR-JST CREST project on Formal Analysis and Design of AI-intensive Cyber-Physical Systems (CyPhAI) funded by the French National Research Agency ANR and Japan Science and Technology Agency JST.






\bibliography{biblio}
\bibliographystyle{plain}

\appendix

\section{\PLstar}\label{apd:plstar}

Here we present \PLstar, a variant of \WLstar~\cite{weiss_WFA_learning}. Similarly, it uses an observation table $O_{\Pref,\Suff}$ for storing outcomes of $\MQ_{\PLstar}$, where $\Pref \subset \Symb^\ast$ is the set of \emph{prefixes} (stored in row indices) and $\Suff \subset \{\terminal\}\cup\Symb^{+}$ is the set of \emph{suffixes} (stored in column indices). 
For every 
$p\in\Pref$ and $s\in\Suff$, $O_{\Pref,\Suff}[p,s] = \MQ_{\PLstar}(p s) = \policyL_{\staI}(p s)$, where $\policyL$ is defined in (\ref{eq:policyL}).
Unlike \WLstar, \Pref\ is divided in two parts~\cite{De_la_higuera:2010}, namely \emph{RED} which are the rows used to construct \emph{states}, and \emph{BLUE} which are the rows representing continuations. The algorithm ensures that for every row in \emph{RED} all its continuations are in \Pref.
\PLstar\ consists of three main steps. The first one expands $O_{\Pref,\Suff}$ through the use of $\MQ_{\PLstar}$\ until it becomes \emph{closed} and \emph{consistent}. The second one constructs a hypothesis automaton using a greedy clustering technique rather than DBScan. The third one calls $\EQ_{\PLstar}$\ with the proposed hypothesis. For this we adapted Hopcroft-Karp algorithm for checking \tol-similarity. When \Aut\ and $\widehat{\Aut}$ are found to be not \tol-similar, $\EQ_{\PLstar}$ returns a counterexample which is added to \emph{RED} together with all its prefixes, and all their continuations to  \emph{BLUE} (provided they are not already in \emph{RED}). These steps are repeated as long as $\EQ_{\PLstar}$\ yields a counterexample, otherwise it stops and returns the last hypothesis.

%

\section{Proofs}

\paragraph{Proof of Proposition \ref{prop:congruence}}
\begin{enumerate}
    \item By hypothesis, $\policyW(\emptyW | \sta_1) = \policyW(\emptyW | \sta_2)$. By definition, $\policyW(\emptyW | \sta_i) = \policy(\sta_i)$, $i=1,2$. Therefore, $\policy(\sta_1) = \policy(\sta_2)$.
    \item By hypothesis, $\forall\wor\in\Words$, $\policyW(\symb\wor|\sta_1) = \policyW(\symb\wor|\sta_2)$.  By definition, $\policyW(\symb\wor|\sta_i) = \policyW(\wor|\tra(\sta_i,\symb))$, $i=1,2$. Thus, $\forall\wor\in\Words$, $\policyW(\wor|\tra(\sta_1,\symb)) = \policyW(\wor|\tra(\sta_2,\symb))$. Hence, $\tra(\sta_1, \symb)\indist\tra(\sta_2, \symb)$. \qedsymbol
\end{enumerate}

\paragraph{Proof of Proposition \ref{prop:quotient}}
Clearly, for all $\sta, \sta' \in \Sta$, if $\sta\indist\sta'$ then $\probP(\wor|\sta) = \probP(\wor|\sta')$ for every $\wor\in\Symb^\ast$. Hence, $\probP(\wor|\cla{\staI}) = \probP(\wor|\staI)$. \qedsymbol

\paragraph{Proof of Proposition \ref{prop:wm}}
Suppose that \Aut~is not weakly minimal. Then, there are states in \Aut\ which are equivalent. Thus, $\qindist{\Aut}$, which computes the same function as \Aut, has strictly less states than \Aut, which contradicts the hypothesis. \qedsymbol

\paragraph{Proof of Proposition \ref{prop:quantp_vs_tol}}
$\sta\indist_\quantp\sta'$ implies $\policyW(\wor|\sta) =_\quantp \policyW(\wor|\sta')$ for all $\wor\in\Symb^\ast$. Therefore, $\policyL_\sta(\wor\symb) =_\quantp \policyL_{\sta'}(\wor\symb)$  for all $\wor\in\Symb^\ast, \symb\in\SymbT$. Then, $|\policyL_\sta(\wor\symb)-\policyL_{\sta'}(\wor\symb)| \leq \quantp^{-1}$ for all $\wor\in\Symb^\ast, \symb\in\SymbT$. Hence, $\sta\tequal_{\quantp^{-1}}\sta'$. \qedsymbol

\paragraph{Proof of Proposition \ref{prop:tree-size}}
If $\wor, \wor' \in \AccS$ then $\cla{\wor}\neq\cla{\wor'}$, hence $|\AccS| \leq |\qindistp{\Sta}|$. \qedsymbol

\paragraph{Proof of Proposition \ref{prop:termination}}~\\
\noindent
\textit{Termination} First notice that all inner operations terminate. Second, when a counterexample is returned, the number of leaves of \NTree\ increases by at least 1. Therefore by Proposition \ref{prop:tree-size} and the fact that $|\qindistp{\Sta}| \leq |\Sta|$, $\QuaNT$ terminates.\\ 
\noindent
\textit{Correctness} When $\QuaNT$ terminates, the hypothesis $\widehat{\Aut}$ constructed is such that $\widehat{\Sta} = \qindistp{\Sta}$, $\widehat{\policy} = \qindistp{\policy}$, $\widehat{\tra} =  \qindistp{\tra}$. Therefore $\widehat{\Aut} \indist_\quantp \Aut$. \qedsymbol

\section{Figures}
\

\begin{figure}[htbp]
\centering
    \begin{subfigure}[c]{0.45\textwidth}
    \begin{tikzpicture}
        \node[state, initial, initial where=above] (q0) {\stackanchor{$q_0$}{0}};
        \node[state, below of=q0] (q1) {\stackanchor{$q_1$}{0}};
        \node[state, right of=q1] (q2) {\stackanchor{$q_2$}{0.3}};
        \node[state, below of=q1] (q3) {\stackanchor{$q_3$}{0}};
        \draw   
            (q0) edge[left] node{$a/0.1$} (q1)
            (q0) edge[below, bend right=75, left] node{$b/0.9$} (q3)
            (q1) edge[below] node{$b/0.2$} (q2)
            (q1) edge[left] node{$a/0.8$} (q3)
            (q2) edge[bend right, above] node{$a/0.1$} (q1)
            (q2) edge[below, bend left, right] node{$b/0.6$} (q3)
            (q3) edge[loop below] node{$a,b/0.5$} (q3);
    \end{tikzpicture}
    \end{subfigure}
    \begin{subfigure}[c]{0.45\textwidth}
    \begin{tikzpicture}
        \node[state, initial, initial where=above] (q0) {\stackanchor{$q'_0$}{0}};
        \node[state, below of=q0] (q1) {\stackanchor{$q'_1$}{0}};
        \node[state, right of=q1] (q2) {\stackanchor{$q'_2$}{0.3}};
        \node[state, below of=q1] (q3) {\stackanchor{$q'_3$}{0}};
        \node[state, above of=q2] (q4) {\stackanchor{$q'_4$}{0}};
        \draw   
            (q0) edge[left] node{$a/0.1$} (q1)
            (q0) edge[below, bend right=75, left] node{$b/0.9$} (q3)
            (q1) edge[below] node{$b/0.2$} (q2)
            (q1) edge[left] node{$a/0.8$} (q3)
            (q2) edge[right] node{$a/0.2$} (q4)
            (q2) edge[below, right] node{$b/0.5$} (q3)
            (q3) edge[loop below] node{$a,b/0.5$} (q3)
            (q4) edge[bend right, left] node{$b/0.1$} (q2)
            (q4) edge[right, bend left=90] node{$a/0.9$} (q3);
    \end{tikzpicture}
    \end{subfigure}
    \caption{PDFA $A$ (left) and $B$ (right).}
    \label{fig:pdfa_A_B}
\end{figure}
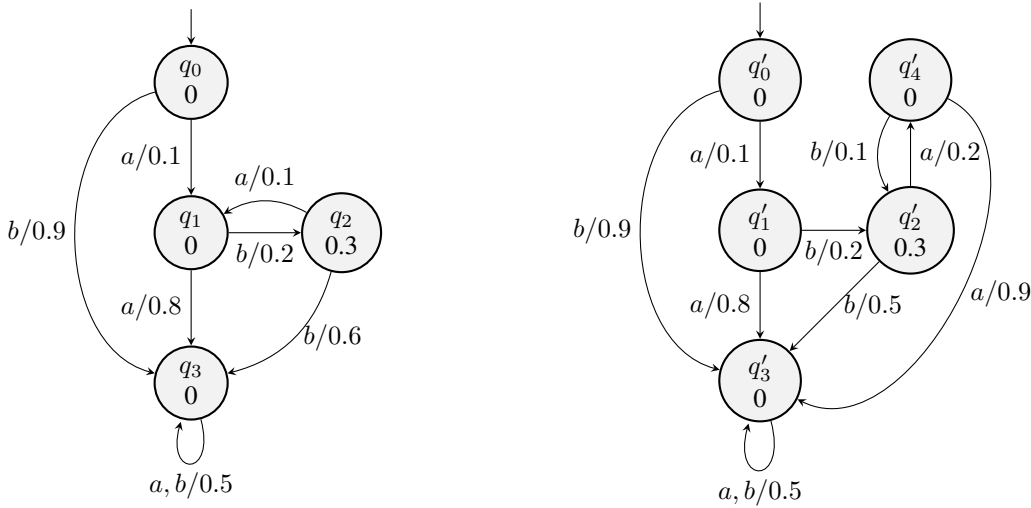

\end{document}